\documentclass[useAMS,usenatbib]{mn2e}
\usepackage{graphicx}

\title{Cosmological black holes as voids progenitors. I. Simulations.}
\author[M. Serpico]
       {M. Serpico$^1$, R. D'Abrusco$^1$, G. Longo$^{1,2,3}$, C. Stornaiolo$^{1,2}$ \\
        1 - Department of Physical Sciences, University Federico II of Napoli, ITALY\\
        2 - INFN - Napoli Unit, via Cinthia 6, 80100 ITALY\\
        3 - INAF - Napoli Unit, via Moiariello 16, 80131 Napoli, ITALY}

\date{Accepted xxxxxxx; received xxxxxxx; in original form 2005 July 12}

\pagerange{\pageref{firstpage}--\pageref{lastpage}}
\pubyear{1994}

\begin{document}

\maketitle

\label{firstpage}

\begin{abstract}
Cosmological black holes (CBH), {\it i.e.} black holes with masses
larger than $10^{14} \ M_{\odot}$, have been proposed as possible
progenitors of galaxy voids~\cite{Stornaiolo_a}. The presence of a
CBH in the central regions of a void should induce significant
gravitational lensing effects and in this paper we discuss such
gravitational signatures using simulated data. These signatures
may be summarized as follows: i) a blind spot in the projected
position of the CBH where no objects can be detected; ii) an
excess of faint secondary images;  iii) an excess of double images
having a characteristic angular separation. All these signatures
are shown to be detectable in future deep surveys.
\end{abstract}


\section{Introduction}\label{introduction}
Voids are among the largest structures known in the Universe with
typical diameters ranging between 20 and 85 Mpc. Among the various
models so far proposed for their formation, we may quote the one
originally introduced in \cite{Friedmann}, accordingly to which
voids form from the evolution of negative primordial perturbations
in the density field. More in detail, in this model void formation
is the result of two correlated processes. The first one is the
comoving expansion of these negative fluctuations. The second
arises from the biased galaxy formation picture: galaxies are less
likely to form in the underdense regions created by this
expansion. Several N-body simulations of this formation mechanism
based on the cold dark matter scenario (cf. \cite{Benson})
produced results which are consistent with observational data.
However, it needs to be stressed that the current observational
samples of data (void surveys, morphological classification of
void galaxies) \cite{Rojas} are too small to allow any conclusive
test of this or other models.

Another formation mechanism was proposed by Stornaiolo
\cite{Stornaiolo_a}. In this scenario the collapse of extremely
large wavelength positive perturbations led to the formation of
low density/high mass black holes (Cosmological Black Holes or
CBH). Voids are then formed by the comoving expansion of the
matter surrounding the collapsed perturbation. This model implies
that at the center of voids should still exist a very massive ($M
\,>\, 10^{14}\,M_{\odot}$) CBH, which should be detectable through
the gravitational lensing effects induced on background galaxies.
In this paper we discuss what these effects are and whether they
might be actually observable or not.

This paper is structured as follows. In Section~\ref{simulation}
we present the simulations which were performed in order to derive
the possible gravitational signatures of the CBH, summarized in
Section~\ref{observquant}. In Section~\ref{observations} we
discuss whether such effects may or may not be observed in
existing or ongoing surveys. In Section~\ref{conclusions} we draw
some conclusions. In a subsequent paper we shall discuss the weak
lensing effects possibly induced by a CBH on a background galaxy
distribution.

\section{The simulations}\label{simulation}

In order to evaluate the gravitational lensing effects induced by
a CBH located in the center of a void we produced two sets of
simulations: one set for a reference unperturbed universe, {\it
i.e.} for a void in an otherwise uniform universe (without the
CBH) and a second set obtained from the previous by adding a CBH
in the center of the void.

In both cases we assumed, for simplicity, an Einstein-de Sitter
cosmological background model, with $\Omega_M=1$, $H_{0}=100h\,\,
Km\,s^{-1}\,Mpc^{-1}$ and no cosmological constant. We wish to
stress that the only effect of a non zero cosmological constant
would be to change the relation
$D=D(z,H_{0},\Omega_{M},\Omega_{\Lambda})$ between the distance
$D$ and the redshift $z$ of the background galaxies which is
involved in the calculation of the apparent magnitudes and of the
angular positions. Using approximate expressions \cite{Kantowski}
it is easy to see that over a redshift range $\left[ 0,0.4\right]$
the maximum deviation from the standard Einstein-de Sitter model
would be of the order of $10\%$.

As mentioned, the reference universe is described by an
Einstein-de Sitter background containing a spherical void of
radius $R_{void}$ located at the comoving distance $D_{void}$ from
the observer which does not produces any detectable deflection or
magnification effects on background galaxies (as pointed out in
\cite{Amendola}).

The perturbed universe is instead described by a Swiss-Cheese
model, with a CBH in the center of the void, which, as above, has
comoving radius $R_{void}$ and is located at the comoving distance
$D_{void}$ from the observer. The CBH has a mass
\begin{equation}
M=\frac{4}{3}\,\pi\,\Omega_{cbh}\,\rho_{crit}R^{3}_{void}
\end{equation}
where $\Omega_{cbh}$ is the density parameter of the CBH.
\begin{figure}\label{CBH-schema1}
\centerline{\includegraphics[width=6cm]{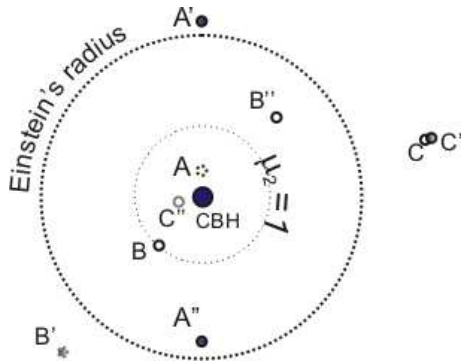}}
\caption{Schematic layout of how the CBH gravitational effects
affect the formation of double images as a function of the angular
distance of the source from the CBH (see text for a more detailed
explanation).}
\end{figure}
It needs to be stressed that while the background universe is
described by a FLRW metric, inside the void region holds a
Schwarzschild metric. The boundary conditions at the transition
between the two regimes are discussed in \cite{Kantowski69}. We
also assume that changes in the sources redshift, due to the
propagation of light across the edge of the void, is negligible.
In other words, we describe the void-CBH system as a Schwarzschild
lens enclosed in a Einstein-de Sitter cosmological model. \\
More
in particular we simulated a slice of universe covering a solid
angle defined by the void diameter and a redshift range comprised
between the far edge of the void and $z=0.4$. This conical volume
was populated with a randomly distributed (in the comoving frame)
galaxy population drawn from the $r$-band SDSS (Sloan Digital Sky
Survey) luminosity function for the field \cite{Blanton}.
Evolutionary effects induced by the redshift on the luminosity
function were neglected. Furthermore we treated galaxies as
material points since their physical extension is not relevant to
the following discussion.

The lensing effects induced by the CBH were then derived in the
weak lensing approximation \cite{Schneider} by assuming that the
light from the sources passed at a distance from the CBH much
larger than its Schwarzschild radius (about $30\,arcsec$ for
$M=10^{14} \ M_{\odot}$ and $D_{void}=50\,Mpc$).

The deflection angle produced by the CBH for radiation approaching
with impact parameter $\xi$ is given by
\begin{equation}
\hat{\alpha}=\frac{4GM}{c^{2}\xi}
\end{equation}

For a source at distance $D_{s}$ from the observer, the Einstein
angle can be written in the form:
\begin{equation}
\hat{\alpha}_{0}=\sqrt{\frac{4GM}{c^{2}}\frac{D_{ds}}{D_{void}D_{s}}}
\end{equation}
\noindent where $D_{ds}$ is the distance of the source from the
lens.

If $\hat{\beta}$ is the unlensed angular position of the source
with respect to the observer, we know that the effect of the lens
will be the creation of two images of the source with angular
positions:

\begin{equation}\label{tetaschw}
\hat{\theta}_{1,2}=\frac{1}{2}\left(\hat{\beta}\pm\sqrt{4\hat{\alpha}_{0}^{2}+\hat{\beta}^{2}}\right)
\end{equation}

\noindent and with magnifications given by:

\begin{equation}\label{muschw}
\mu_{1,2}=\frac{1}{4}\left(\frac{\tilde{\beta}}{\sqrt{\tilde{\beta}^{2}+4}}+\frac{\sqrt{\tilde{\beta}^{2}+4}}{\tilde{\beta}}\pm2\right)
\end{equation}

\noindent where $\tilde{\beta}=\hat{\beta}/\hat{\alpha}_{0}$.

Simulations were then performed for different values of
$\Omega_{cbh}$ and $R_{void}$ (0.05 to 0.4 with step 0.05 and from
10 to 20 Mpc, step 2, respectively).

\section{Qualitative description of observable quantities}\label{observquant}
The simulations showed that the CBH leaves three different types
of signatures on the background galaxy distribution. In order to
better quantify what happens we refer to Fig.~1 which shows the
images produced in three different relative positions of source
and lens. When the ``real'' object (A) is very close to the CBH,
we have the formation of two images, namely $A'$ and $A''$ which
are respectively outside and inside the Einstein Radius and very
close to it. Both images are brighter than the unlensed image and
$A'$ is brighter than $A''$. Then when the source moves away from
the CBH, the amplification factor with respect to the secondary
image tends to $1$.

\noindent The dashed inner circle marks the position of such
locus. If the source $B$ lays on this circle, then it will produce
two images $B'$ and $B''$. $B'$ is outside of the Einstein Radius
at a larger distance than $A'$ and is brighter than $B$, while
$B''$ lays inside the Einstein Radius closer to the CBH than
$A''$.

\noindent Finally let us consider the case of a source $C$ which,
if unperturbed would fall outside of the Einstein Radius. Also in
this case we shall see two images $C'$ and $C''$. The first one
will almost coincide with the position of $C$ and will have an
almost identical brightness, while $C''$ will be almost invisible
and very close to the CBH.

When a random distribution of background galaxies is considered,
the overall effect of the lensing will be the formation of 4
different areas on the sky, namely the regions $A$, $B$, $C$ and
$D$, shown in Fig.~2.

\noindent The inner circle A, which we call 'blind spot', is the
region where no image can be detected due to the increasing
demagnification of secondary images when they move towards the
center of the void. The size of the blind spot depends on the mass
of the CBH and on the density of the background galaxy
distribution.

\noindent The annulus B is characterized by the presence of a
large number of secondary images, and it is the second observable
feature associated with the CBH.

\noindent The third zone C, is an annular zone that we call the
'deficit zone'. It encompasses the average Einstein Radius of the
galaxy sample and is characterized by a relatively low number of
background galaxy images. It can be understood reminding that, at
this angular distance from the center of the void, one can find
only those primary and secondary images which originates from
sources having angular position falling well within the Einstein
Radius. Therefore, in this annulus it is expected to find a lack
of both primary and secondary images: as the source moves away
from the Einstein Radius, the primary image tends to coincide with
the original position of the source while the secondary image
becomes fainter and moves more and more towards the CBH.

\noindent Finally, the D zone does not present any particular
feature produced by lensing, and coincides with the homogeneous
background galaxy distribution.

In total we run 48 sets of simulations assuming the void distance
at $50 \ Mpc$ ({\it i.e.} matching the distance of the nearest
void) and covering a grid defined by: $\Omega_{CBH}=0.05
\rightarrow 0.4$ with step $0.05$, and $R_{void}=10 \rightarrow 20
\ Mpc$ with step $2$ Mpc. For each grid point the procedure was
iterated 1000 times randomly changing at each iteration the
positions of the background galaxies. Each simulation produced a
catalogue of galaxy positions and magnitudes and each group of
simulations was then used to derive average quantities.

\section{Results}\label{observations}

The main observational signatures left by the CBH can be
summarized as follows.

\subsection{Blind spot}

For every simulation, we determined an estimate of the angular
radius $\theta_{blind}$ of the blind spot, defined as the minimum
angular distance from the CBH of the secondary images (brighter
than $m_{lim}=23.5$) and then, at each simulation grid point we
took the average value over the 1000 simulations. In Table~1 we
list some representative values.

\begin{table}\label{table_bs}
\begin{center}
\begin{tabular}{l|rrr}
\hline \hline
 $\Omega_{CBH}/R_{void}$&10 &12 &20 \\
 \hline
 0.05& 16$\pm 5$&25$\pm 7$&36$\pm 8$\\
 0.2&33$\pm 8$&36$\pm 9$&53$\pm 10$\\
 0.4&39$\pm 8$&44$\pm 9$&69$\pm 12$\\
 \hline
 \end{tabular}
 \end{center}
 \caption{Average value of $\theta_{blind}$ (in arcsec) as a function of $\Omega_{CBH}$ and $R_{void}$ (in Mpc). The quoted errors are the
 r.m.s. of the 1000 individual simulations. }
\end{table}

As expected, the size of the blind spot, increases with
$\Omega_{CBH}$ and with $R_{void}$. Assuming, for instance, a
typical value of $30 \ arcsec$ (cf. Table~1), it is apparent that
the blind spot should be rather difficult to observe. In fact,
while it should be easily detectable in very deep number counts,
its size is of the same order of the average angular separation
between bright galaxies at intermediate redshift. Possible effects
connected with the distortion induced on extended background
objects which happen to fall near the line of sight of the CBH
will be addressed in a forthcoming paper.
\begin{figure}\label{CBH-schema2}
\centerline{\includegraphics[width=6cm]{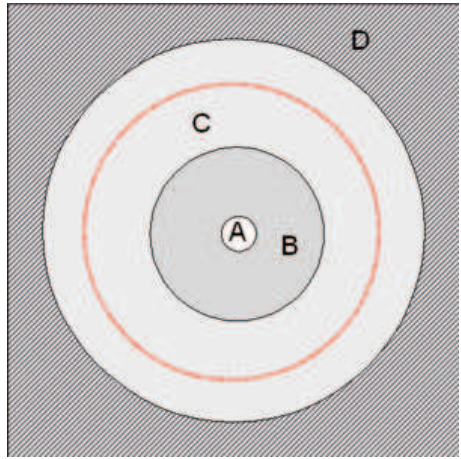}}
\caption{The three regions described in the text. Region A
(greatly enlarged to make it visible) is the blind spot; Region B:
region where we expect the excess of secondary images; Region C:
deficit region (the circle inside the C region shows the average
Einstein Radius).}
\end{figure}
\subsection{Galaxy number counts and radial profile}
\begin{figure}\label{conteggi_1}
\centerline{\includegraphics[width=7cm]{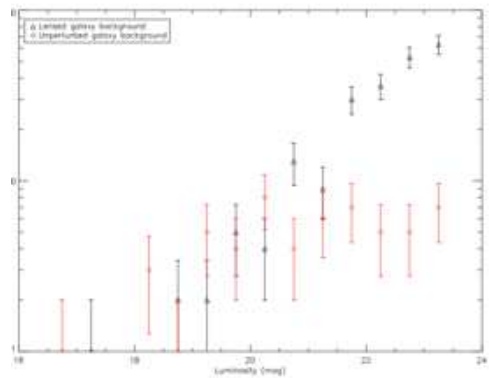}}
\caption{The figure shows in red the average number counts as a
function of the apparent magnitude in the $r$ band, obtained,
respectively, in the outer region (see text) and in black those
obtained in the inner region.}
\end{figure}
The presence of the CBH affects the number counts. In order to
estimate the size of such effect and in absence of a priori
information on the size of the void we adopted the following
procedure. First we introduced an annular zone (defining an inner
and outer region) centered on the blind spot. The radius of the
zone was then found by maximizing the difference between the
average galaxy counts in the inner and outer regions.

As it can be seen in Fig.~3, the average number counts associated
with the inner area show a systematic difference with respect to
those extracted from the background. This effect however becomes
significant only at magnitudes fainter than $\sim 21.0$.
\medskip

In order to quantify such difference we performed a
Kolmogorov-Smirnov test on both distributions of points. Each data
set was split into two parts including galaxies brighter or
fainter than 21 mag, respectively. The brighter parts of the
distribution do not present any statistically significant
difference, while for the fainter parts we derived a probability
higher than $98\%$, that the two samples are drawn from different
populations.
\medskip

As it was discussed in the previous paragraphs, the presence of a
CBH induces a typical pattern in the number counts radial
profiles. Such pattern is characterized by a peak in the range of
distances intermediate between the blind spot radius and the
average Einstein Radius (caused by the secondary images
concentration), followed by a dip, which corresponds to a slight
underdensity of objects and than at distances comparable with the
Einstein Radius it raises up again to smoothly reach the value
expected for the background galaxy distribution.

In Fig.~4a we show the number counts profile extracted from the
simulation grid--point at $\Omega_{CBH}=0.2$ and $R_{void}=12$
Mpc. The first point of the profile is located in the blind spot
and is followed by an isolated peak which rapidly falls in the
dip.
\begin{figure}\label{radial_1}
\centerline{\includegraphics[width=7cm]{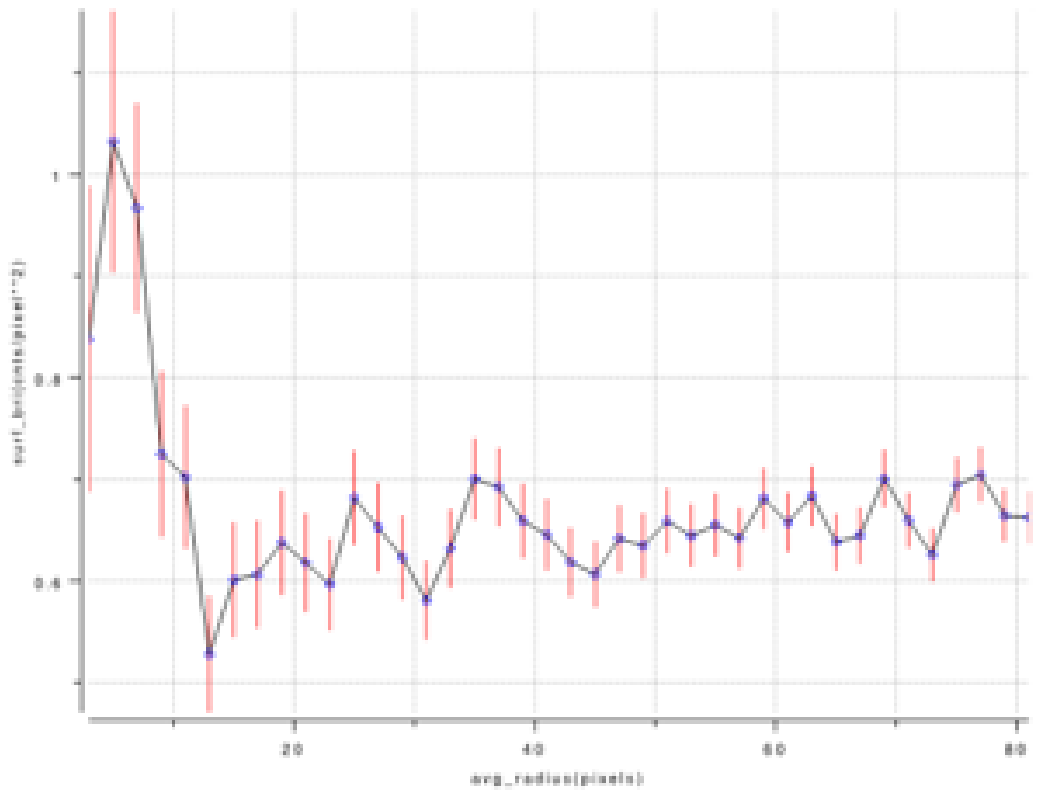}(a)}
\centerline{\includegraphics[width=7cm]{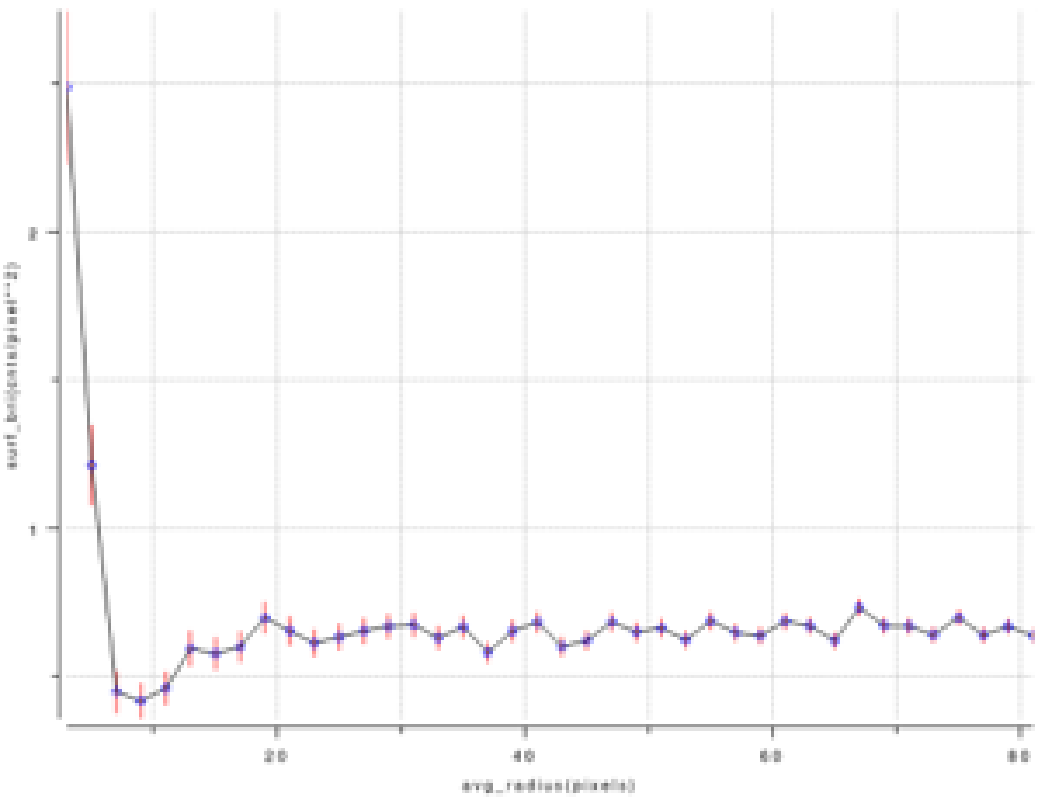}(b)}
\centerline{\includegraphics[width=7cm]{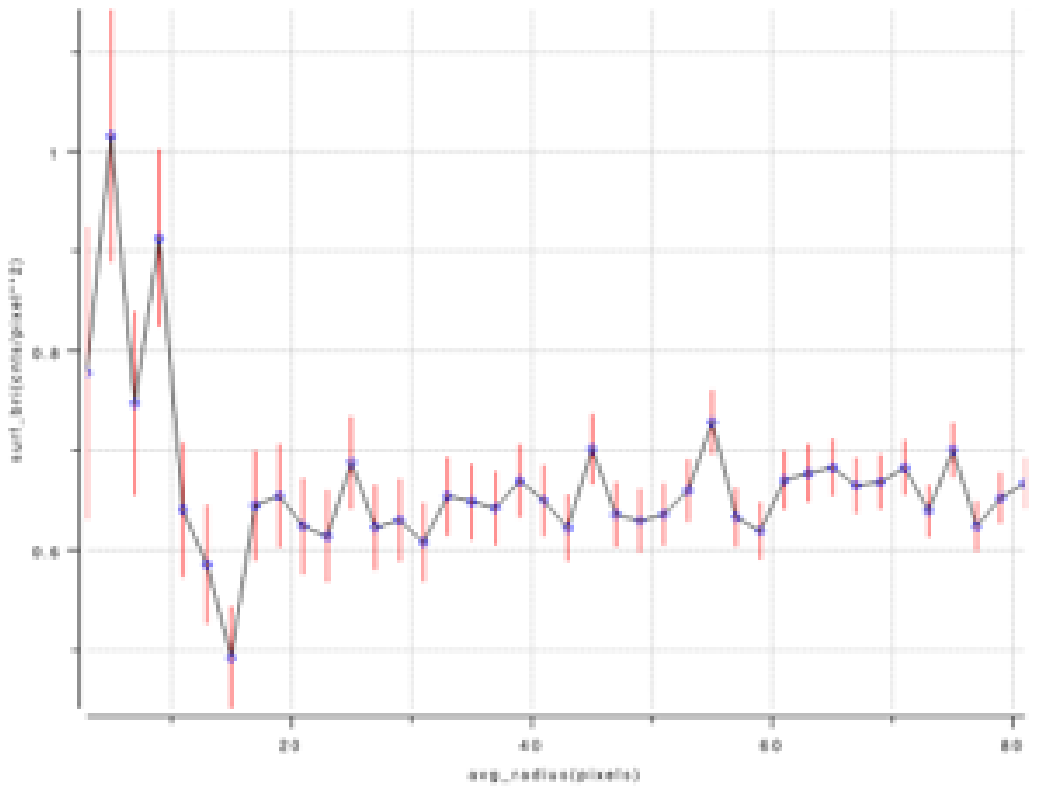}(c)}

\caption{Galaxy number counts radial profiles obtained by
integrating over annular regions centered in different regions.
Panel a - centered on the blind spot (zone A); Panel b - center
falls in the secondary images overdensity region (zone B); Panel c
- center falls in the deficit region (zone C). }
\end{figure}

The only two other examples of possible radial profiles obtained
when not placed on the center of the void are illustrated in
Figg.~4b and 4c. The first image, which can be roughly described
by an initial peak, larger than the previous case, located at the
first steps of the profiles followed by the dip, is the typical
pattern generated when the central point of the radial profile is
positioned inside the secondary images overdensity; the second
profile, characterized by the presence of an initial low spot
followed by two distinct peaks and the dip, is reproduced when the
center of the radial density profile lays out of the circular
overdensity created by the secondary images, and inside the
deficit annulus.
\subsection{Multiple images}
The third signature comes just from the pairs of double images
produced by the CBH. We expect angular separations for these pairs
of the order of two times the average Einstein's angle of the
sample. This means angular separations of the order of some
arcminutes. This forecast is particularly striking because if
these doubles really exist, the only way to observe them
practically is to have an estimate of their wide angular
separation. We measured the distribution of the angular separation
between the double images produced by the CBH obtaining the two
points angular correlation function $w(\theta)$ for the simulated
galaxy distribution. A standard Landy--Szalay \cite{Landy}
estimator was used:
\begin{equation}\label{SL_ind}
w(\theta) = \frac{\langle DD \rangle + \langle RR \rangle -
2\langle DR \rangle}{\langle RR\rangle}
\end{equation}
where $\langle DD \rangle$, $\langle DR \rangle$, $\langle RR
\rangle$, are pair counts in bins of $\theta \pm \delta \theta$
of: data--data, data--random and random--random points,
respectively. The statistic has been demonstrated to be close to a
minimum of variance estimator and to be robust with respect to the
number of random points \cite{Kerscher}. In Fig.~5 we show the
above defined correlation function obtained for the simulation
grid-point at $\Omega_{CBH}=0.2$ and $R_{void}=12$ Mpc; a well
defined peak corresponding to an angular distance of $6$ arcmin,
comparable with the average diameter of the secondary images
overdensity region (Zone B), is visible. Moreover, a slight
anticorrelation is found at distance greater than $10$ arcmin,
while for a random distribution of points no correlation of any
sort should be detected. It needs to be stressed that the peak
observed at very short angular separations is an artifact produced
by the fact that in our simulations galaxies are assumed to be
point-like and would disappear if galaxies were approximated with
extended objects.

\begin{figure}\label{correlation}
\centerline{\includegraphics[width=7cm]{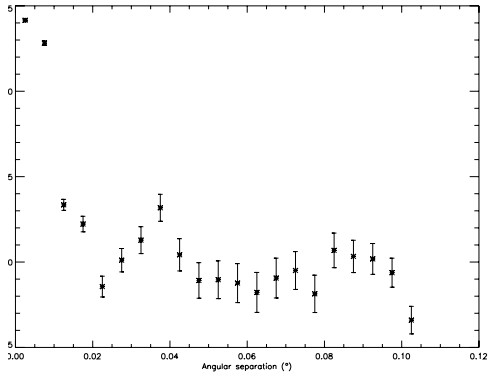}}
\caption{The angular 2--point correlation function derived from
the simulated galaxy distribution (see text).}
\end{figure}

\section{Discussion and conclusion}\label{conclusions}

The results of our simulations which were based on a minimal set
of assumptions and can therefore be regarded as quite general and
robust in their final predictions, clearly show that the presence
of a CBH close to the center of a void would leave unmistakable
signatures on the background galaxy distribution as a result of
the gravitational lensing properties of the CBH. Unfortunately,
they also show that such signatures can be detected only at faint
light levels, {\it i.e.} at magnitudes fainter than the
completeness limit of most existing photometric surveys which
include voids in their field. Furthermore, the only survey which,
at least in theory, should be deep enough to allow at least the
radial profile test above described (namely the Sloan Digital Sky
Survey, cf. \cite{Stoughton}) does not satisfactorily cover any
previously known void. It needs to be stressed however that our
results clearly show that such tests will be possible on any of
the planned deep digital surveys which will become available in
the near future (cf. for instance the VST extragalactic survey).


\begin{thebibliography}{99}
\bibitem[Amendola et al 1999]{Amendola} Amendola L., Frieman
J.A. \& Waga L., 1999, MNRAS, 309, 46
\bibitem[Benson et al 2003]{Benson} Benson A.J., Hoyle F., Torres F. \& Vogeley M.S.,
2003, MNRAS, 340, 160
\bibitem[Blanton et al 2003]{Blanton} Blanton M.R., Hogg D.W., Bahcall N.A., Brinkmann J., et al., 2003,
ApJ, 592, 819
\bibitem[Friedmann and Piran 2001]{Friedmann} Friedmann Y. \& Piran T., 2001, ApJ, 548, 1
\bibitem[Kantowski 1969]{Kantowski69} Kantowski R., 1969, ApJ,
155, 89
\bibitem[Kantowski and Thomas 2001]{Kantowski} Kantowski R. \& Thomas
R.C., 2001, ApJ, 561, 491 5
\bibitem[Kerscher et al 2000]{Kerscher} Kerscher M., Szapudi I.,
Szalay A.S., 2000, ApJ, 335, L13
\bibitem[Landy and Szalay 1993]{Landy} Landy S.D. \& Szalay A.S.,
1993, ApJ, 412, 64
\bibitem[Rojas et al 2005]{Rojas} Rojas R.R., Vogeley M.S., Hoyle
F. \& Brinkmann J., 2005, ApJ, 624, 571
\bibitem[Schneider et al 1992]{Schneider} Schneider P., Ehlers J.
\& Falco E.E. {\it Gravitational Lenses}, Springer
\bibitem[Stornaiolo 2002]{Stornaiolo_a} Stornaiolo C., 2002, Gen.
Rel. \& Grav., 34, 2089
\bibitem[Stoughton et al 2002]{Stoughton}Stoughton C., Lupton R.H., Bernardi M.,
Blanton M.R., et al., 2002, AJ, 123, 485
\end{thebibliography}
\end{document}